\documentclass[aps,prl,reprint,amssymb,amsmath,superscriptaddress]{revtex4-2}
\pdfoutput=1 
\usepackage{graphicx}
\usepackage{dcolumn}
\usepackage{bm}
\usepackage{subfigure}

\begin{document}

\title{Unbalanced clustering and solitary states in coupled excitable systems}%

\author{Igor Franovi\'c}
\email{franovic@ipb.ac.rs}
\affiliation{Scientific Computing Laboratory, Center for the Study of Complex Systems,
Institute of Physics Belgrade, University of Belgrade, Pregrevica 118, 11080 Belgrade, Serbia}
\author{Sebastian Eydam}
\email{sebastian.eydam@gmail.com}
\affiliation{Neural Circuits and Computations Unit, RIKEN Center for Brain Science, 2-1 Hirosawa, 351-0106 Wako, Japan}
\author{Nadezhda Semenova}
\email{nadya.i.semenova@gmail.com}
\affiliation{Saratov State University, Astrakhanskaya str. 83, Saratov, 410012 Russia \\
D\'{e}partment d'Optique P. M. Duffieux, Institut FEMTO-ST, Universit\'{e} Bourgogne-Franche-Comte CNRS UMR 6174, Besan\c{c}on, France}
\author{Anna Zakharova}
\email{anna.zakharova@tu-berlin.de}
\affiliation{Institut f\"{u}r Theoretische Physik, Technische Universit\"{a}t Berlin, Hardenbergstr. 36, 10623 Berlin, Germany}

\date{\today}

\begin{abstract}
We demonstrate the mechanisms of emergence and the link between two types of symmetry-broken states,
the unbalanced periodic two-cluster states and solitary states, in coupled \emph{excitable} systems with prevalent repulsive interactions. Solitary states in non-locally coupled arrays inherit their dynamical features from unbalanced cluster states in globally coupled networks. Apart from self-organization based on phase-locked synchrony, interplay of excitability and local multiscale dynamics also gives rise to \emph{leap-frog} states involving alternating order of spiking. Noise suppresses multistability of cluster states and induces pattern homogenization, transforming solitary states into patterns of patched synchrony.
\end{abstract}


\maketitle

The remarkable discovery of chimera states \cite{KB02,AS04} has led to a profound change of paradigm in understanding of self-organization in assemblies of coupled oscillators. Instead of synchronization transition and onset of collective mode \cite{PRK03}, attention has shifted to states emerging via symmetry breaking of synchrony \cite{M10}, where assemblies of indistinguishable oscillators with symmetric couplings split into groups with different dynamics. A common ingredient to many systems displaying clustering \cite{HMM93,KGO15,KHK19,IA15}, chimeras \cite{Z20,PA15,O18,PJAS21} or solitary states \cite{MPR14,JMK15,JBLDKM18,MSO20,RASZ19,BPSY20} are the long-range interactions. In contrast to low-dimensional dynamics of cluster states, where units within each group behave identically, chimeras are patterns of coexisting coherence and incoherence \cite{Z20}. Similar coexistence underlies solitary states, where a single or a small subset of solitary units split from the synchronized cluster, but unlike chimeras, the minority units spread randomly instead of forming localized domains. Another difference is that solitary states involve spatial chaos \cite{OMHS11}, reflecting sensitive dependence of the dynamics on spatial coordinates, which gives rise to extensive multistability where the number of states grows exponentially with system size. Still, both chimeras and solitary states conform to the class of weak chimeras \cite{AB15}.

For coupled oscillators, much progress has been made in resolving two main problems, namely the mechanisms of the onset and potential links between symmetry-broken states along the path from complete coherence to incoherence. In particular, emergence of clusters from complete synchrony has
been explained by unfolding of cluster singularity, revealing cascade transitions from  synchronous state to balanced two-cluster partition via unbalanced cluster states \cite{KHK19}. Also, clustering has been identified as a prerequisite for the onset of chimeras \cite{SK15}. 
Self-organization of strong chimeras has been shown to involve stabilization of coherent cluster by the incoherent one \cite{ZM21}, while solitary states have been found to mediate transition from complete coherence to chimeras \cite{JMK15}.

However, in a myriad of examples, from neural and cardiac tissue to chemical reactions and lasers,
system components are not intrinsic oscillators, but are rather \emph{excitable} units \cite{LGNS04,I07}, nonlinear threshold elements which in the absence of input lie at rest, but may be triggered to oscillate by sufficiently strong perturbations. There is no reason to expect \emph{a priori} that the results for coupled oscillators translate to realm of excitable systems, where even the onset of collective oscillations requires repulsive rather than attractive interactions. Thus, the fundamental questions on the emergence and the relation between periodic cluster states, chimeras and solitary states in coupled excitable systems remain open.

In this Letter, we address these issues by revealing the mechanisms of onset and the link between different types of unbalanced periodic two-cluster states and solitary states in coupled excitable
systems with multiscale local dynamics, typical for neuroscience \cite{I07}, and varying attractive/repulsive \cite{BRZ15} character of interactions. We show that solitary states in non-locally coupled arrays inherit dynamical features from unbalanced periodic two-cluster states in a globally coupled network. Apart from symmetry-broken states with phase-locked spiking, we discover peculiar cluster and solitary states where self-organization is based on \emph{leap-frog} dynamics \cite{EFW19,GE02,AKW03,OM09}, which involves alternating order of spiking (leader-switching) between units. Leap-frogging cannot exist for phase oscillators, and has recently been shown to emerge from slow-fast dynamics in vicinity of the canard transition \cite{EFW19}. Excitable systems are highly sensitive to noise \cite{LGNS04}, and it has been known to play a facilitatory role for spontaneous clustering \cite{FTVB12} and emergence of chimeras \cite{SZAS16,ZSAS17,SEM20}. Here we demonstrate the effect 
of noise-induced preference of attractors \cite{PF14,K97,PJSHT11}, where the noise suppresses system's multistability by promoting certain types of cluster states as well as patched patterns at the
expense of solitary states.

Our system is an array of $N$ identical FitzHugh-Nagumo units whose dynamics is given by
\begin{align}
\varepsilon \frac{du_k}{dt}&=u_k-\frac{u_k^3}{3}-v_k +\frac{\kappa}{2R}\sum\limits_{l=k-R}^{k+R}
[g_{uu}(u_l-u_k) \nonumber \\
&+g_{uv}(v_l-v_k)] +\sqrt{\varepsilon}\sigma \xi_k(t) \nonumber \\
\frac{dv_k}{dt}&= u_k+ a +\frac{\kappa}{2R}\sum\limits_{l=k-R}^{k+R}
[g_{vu}(u_l-u_k) \nonumber \\
&+g_{vv}(v_l-v_k)], \label{eq:csys}
\end{align}
where the local slow-fast dynamics is governed by the activator variables $u_k$ and the recovery variables $v_k$, and the scale separation is introduced by a small parameter $\varepsilon=0.05$. All indices are modulo $N$. For neuronal systems, fast and slow variables describe the dynamics of membrane potential and the coarse-grained action of gating variables, respectively. Local bifurcation parameter $a$, here fixed to $a=1.001$, mediates the transition from excitable ($|a|>1$) to oscillatory behaviour ($|a|<1$). Due to the singular character of Hopf bifurcation at $a=1$, the onset of oscillations is followed by a \emph{canard transition} ($a\approx 1-\varepsilon/8$) from small-amplitude (subthreshold) to large-amplitude relaxation oscillations \cite{BE86}. Non-local interactions are characterized by the coupling strength $\kappa$, here fixed to $\kappa=0.4$, and the radius $r=R/N$, whereby the case $r=1/2$ conforms to a globally connected network. Apart from terms involving only $u_k$ or $v_k$, there are also cross-coupling terms, which is compactly written via rotational coupling matrix \cite{OOHS13}
$G = \left(\begin{matrix}g_{uu} & g_{uv}\\ g_{vu} & g_{vv}\end{matrix}\right)=
\left(\begin{matrix}\cos \phi & \sin \phi\\ -\sin \phi & \cos \phi\end{matrix}\right)$. Parameter $\phi$ affects the prevalence of attractive and repulsive interactions. The latter are necessary to induce oscillations as attractive interactions cannot destabilize the collective rest state.
Spiking can also emerge due to noise, acting in the fast variables in analogy to synaptic noise \cite{DRL12}. Each unit is influenced by independent Gaussian white noise $\xi_k(t)$ of intensity $\sigma$. We first focus on how the stability of unbalanced two-cluster states in globally connected networks changes with $\phi$ and then analyze the onset of solitary states in non-locally coupled arrays ($r<1/2$) of excitable elements. Previously, solitary states were studied for FitzHugh-Nagumo oscillators in one- \cite{RASZ19,MRJZ19} and two-layer \cite{SJMZ21,RZS21} networks.

\paragraph{Two-cluster states in a globally coupled network} To gain insight into the structure of unbalanced periodic two-cluster states, their stability domains and bifurcations at their boundaries, we implement a twofold approach, combining the semi-analytical method of \emph{evaporation exponents}
and the numerical path-following method based on introducing \emph{probe oscillators}. Since our interest is in solutions where both clusters emit spikes, the splitting scenario by which clusters emerge from the collective rest state is beyond the scope of this Letter. We just note that for stable local dynamics ($|a|>1$), interaction-induced destabilization of the stationary state at $\phi^*=\arccos(\frac{1-a^2}{2\kappa})$ is a highly degenerate point where $2(N-1)$ Jacobian eigenvalues simultaneously become critical, giving rise to a large number of different cluster partitions featuring subthreshold oscillations, which in an exponentially small $\phi$ region start to display spikes via secondary canard transitions. Stability of the stationary state is regained at $\bar{\phi}=\phi^*+\pi$. Onsets of cluster instability and subsequent periodic cluster states for type I excitable units are addressed in \cite{RZ21,ZT16}.

Unlike Lyapunov exponents, evaporation exponents \cite{PP16,ZP17,PPM01} can describe perturbations that destroy cluster partitions. They characterize stability of clusters to emanation of elements, induced by perturbations transversal to invariant subspace of a particular partition. Clusters with negative evaporation exponents are attractors of an assembly, while their positive values indicate instability. Let us consider a two-cluster state with $N_A=pN$ units in cluster A and $N_B=(1-p)N$ units in cluster B. The dynamics of this state is independent on $N$, and is governed by the reduced system
\begin{align}
\varepsilon\dot{u_{i}}&=u_{i}-\frac{1}{3}u_{i}^{3}-v_{i}+\kappa w_i(g_{uu}(u_{j}-u_{i})+g_{uv}(v_{j}-v_{i})) \nonumber \\
\dot{v_{i}}&=u_{i}+a+\kappa w_i(g_{vu}(u_{j}-u_{i})+g_{vv}(v_{j}-v_{i})), \label{eq:reds}
\end{align}
with $i,j\in\{A,B\},i\neq j$, and $(w_A,w_B)=(1-p,p)$ being additional coupling weights derived from particular cluster partition. For $p\neq 1/2$, system \eqref{eq:reds} is equivalent to a pair of nonidentical excitable units with attractive and repulsive interactions. Each value of partition parameter $p$ specifies an invariant subspace in the complete phase space, so that subspaces for different $p$ intersect only in the full synchrony plane $u_A=u_B,v_A=v_B$. To introduce evaporation exponents, we consider symmetric small perturbations to two units, 1 and 2, in each of the clusters: $u_{i,1/2}=u_i \pm \delta u_i, v_{i,1/2}=v_i \pm \delta v_i, i\in\{A,B\}$. Due to permutation symmetry, they can be applied to an arbitrary pair of elements, leaving the cluster mean-fields unchanged. Linearized equations for deviations $(\delta u_{i}(t), \delta v_{i}(t))$ transversal to clusters $i\in\{A,B\}$ read
\begin{align}
\varepsilon\dot{\delta u_{i}}&=(1-u_{i}^{2}-\kappa g_{uu})\delta u_{i}-(1+\kappa g_{uv})\delta v_{i} \nonumber \\
\dot{\delta v_{i}}&=(1-\kappa g_{vu})\delta u_{i}-\kappa g_{vv}\delta v_{i}. \label{eq:dev}
\end{align}
Evaporation exponents $\lambda_{ev,i}=\underset{T\rightarrow\infty}{lim}\frac{1}{2}\ln\frac{\delta u_i^{2}(T)+\delta v_i^{2}(T)}{\delta u_i^{2}(0)+\delta v_i^{2}(0)}$ for $i\in\left\{A,B\right\}$
are obtained by integrating system \eqref{eq:reds}-\eqref{eq:dev}.

\begin{figure}[ht]
\centering
\includegraphics[scale=0.84]{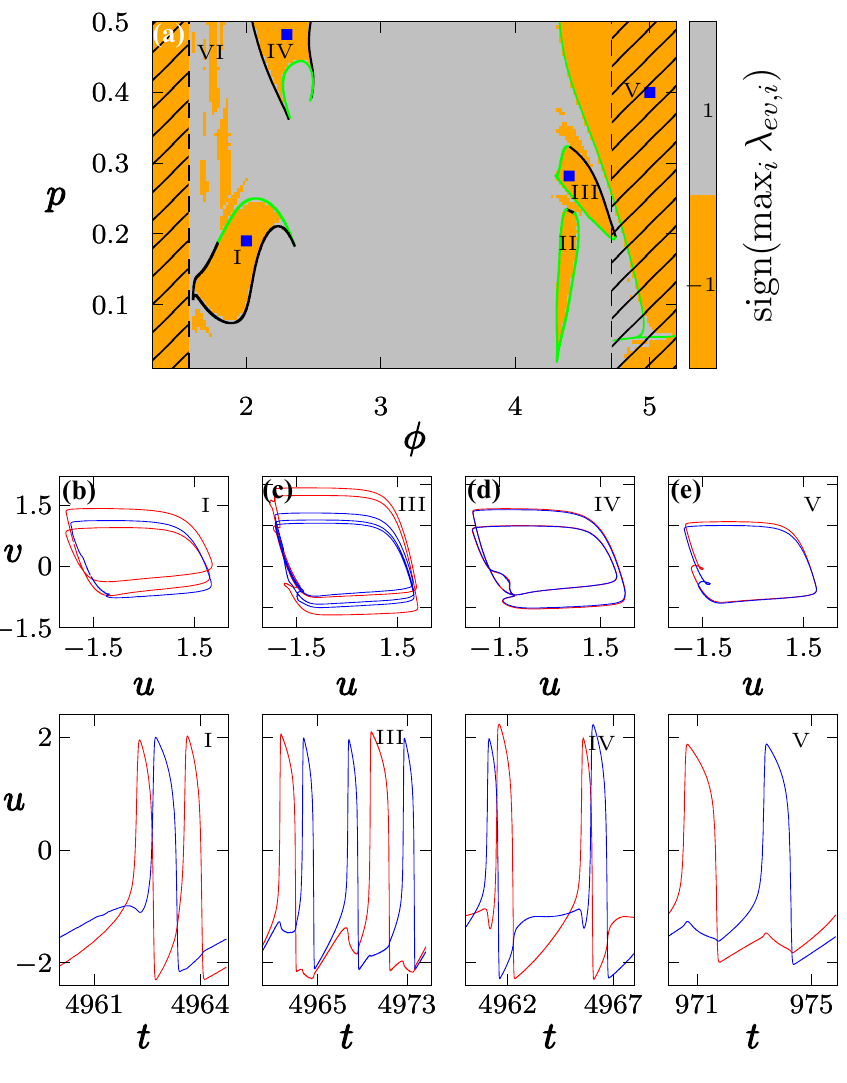}
\vspace{-0.4cm}
\caption{Unbalanced periodic two-cluster states in globally coupled network.(a) Stability diagram in the $(\phi,p)$ parameter plane. Darker (lighter)
shading: stable (unstable) solutions. Black solid lines: period-doubling bifurcations, green lines: curves of branching points. Black dashed lines: destabilization and reappearance of collective rest state at $\phi=\phi^*\approx 1.573$ and $\phi=\bar{\phi}\approx 4.715$. (b)-(e) Time traces $u_i(t),i\in\{A,B\}$ and phase portraits corresponding to characteristic domains from (a). Associated $(\phi,p)$ values are shown by blue squares in (a). Remaining parameters are $\kappa=0.4,a=1.001,\varepsilon=0.05$.}
\label{fig:std}
\vspace{-0.3cm}
\end{figure}
Bifurcations at stability boundaries of particular cluster states are determined by numerical continuation using probe oscillators, designed to test whether a unit added to the cluster asymptotically remains in or emanates from it. 
Probes are introduced to each of the clusters by coupling to the corresponding mean-fields, but without affecting the mean-fields themselves, such that their dynamics $(\tilde{u}_i(t),\tilde{v}_i(t))$ obeys
\begin{align}
\varepsilon \dot{\tilde{u}}_i&=\tilde{u}_i-\tilde{u}_i^3-\tilde{v}_i+\kappa[w_i(g_{uu}(u_i-\tilde{u}_i)+g_{uv}(v_i-\tilde{v}_i)) \nonumber \\
&+w_j(g_{uu}(u_j-\tilde{u}_i)+g_{uv}(v_j-\tilde{v}_i))] \nonumber \\
\dot{\tilde{v}}_i&=\tilde{u}_i+a+\kappa[w_i(g_{vu}(u_i-\tilde{u}_i)+g_{vv}(v_i-\tilde{v}_i)) \nonumber \\
&+w_j(g_{vu}(u_j-\tilde{u}_i)+g_{vu}( v_j-\tilde{v}_i))], \label{eq:probes}
\end{align}
for $i,j\in\{A,B\},i\neq j$. Continuation of system \eqref{eq:reds} together with \eqref{eq:probes} was performed by 
AUTO \cite{AUTO12}.

Combined results of the two approaches shown in Fig.~\ref{fig:std}(a) summarize the stability diagram for the system \eqref{eq:reds}-\eqref{eq:dev} in the $(\phi,p)$ plane, with darker (lighter) shading indicating the stable (unstable) solutions. Black curves at the boundaries of stable regions indicate period-doubling bifurcations, whereas the green lines correspond to curves of branching points. The latter are typically pitchfork bifurcations in the reduced system, but correspond to unfolding of highly degenerate bifurcation points \cite{KHK19} of system \eqref{eq:csys}, where $p$ becomes a solution parameter. System \eqref{eq:reds} supports six characteristic regimes with 1:1 (regions IV, V and VI), 1:2 (I, II) and 2:3 (region III) frequency locking, all conforming to mixed-mode oscillations \cite{DGKKOW12,K15} with interspersed large- and small-amplitude oscillations, cf. Fig.~\ref{fig:std}(b)-(f). Apart from the classical phase-locked solutions I,II,III and V, one also observes unbalanced cluster states with leap-frog patterns \cite{EFW19,GE02,AKW03,OM09} between the clusters. Switching of leadership occurs via subthreshold oscillations, such that the current leader performs an extra small oscillation allowing it to be overtaken by the lagging cluster. Stable
two-cluster solutions at the line $p=1/2$ may acquire an additional antiphase space-time symmetry $u_A(t)=u_B(t+P/2),v_A(t)=v_B(t+P/2)$, where $P$ is the oscillation period. Different types of leap-frog patterns and the underlying mechanisms have recently been analysed in \cite{EFW19}, showing that they may qualitatively be understood within the framework of phase-sensitive excitability of a periodic orbit \cite{FOW18}, a new concept manifested as a strong and non-uniform sensitivity to perturbations of both the small- and large-amplitude oscillations in the FitzHugh-Nagumo system.

\begin{figure}[ht]
\vspace{-0.2cm}
\centering
\includegraphics[scale=0.9]{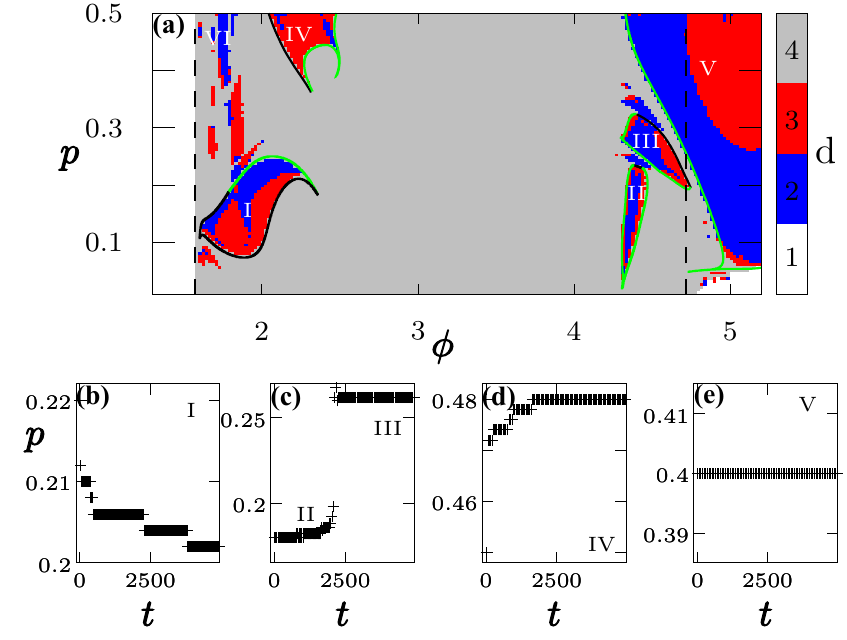}
\vspace{-0.3cm}
\caption{Persistence of unbalanced periodic two-cluster states under noise. (a) Quantity $d(\phi,p)$ distinguishes between four cases: cluster states reorganize to a partition with smaller (blue, $\lambda_{ev,A}>\lambda_{ev,B}$) or larger $p$ values (red, $\lambda_{ev,B}>\lambda_{ev,A}$); two-cluster states are unstable (gray, $d=4$); only the synchronous stationary state is stable (white, $d=1$). (b)-(e) Examples of evolution of partition parameter $p(t)$ under noise. From left to right, interaction parameters and noise levels are $\phi=2,4.4,2.3,5.0$, $\sigma=5,0.6,0.6,5 \times 10^{-3}$, respectively.}\label{fig:ncl}
\vspace{-0.1cm}
\end{figure}
Evaporation exponents can also be used to approximate the impact of sufficiently small noise to stability of two-cluster partitions. In particular, for \eqref{eq:csys} with $r=1/2, \sigma>0$, we find that noise may cause a transition to another type of two-cluster state, or may reorganize the state's structure by inducing migration of units between the clusters, without qualitatively changing the dynamics of cluster mean-fields. Reorganization process eventually settles to a partition where the net transport between the clusters reaches a dynamical balance. Of course, both splitting of a unit from a cluster and migration to another one may involve nonlinear effects. Still, at the linear level, the "potential barrier" that has to be overcome when unit leaves the cluster is proportional to $\lambda_{ev,i}$. This is used to characterize susceptibility of two-cluster states to noise in Fig.~\ref{fig:ncl}(a). We distinguish between the cases where the noise is more likely to shift a two-cluster partition toward smaller ($0>\lambda_{ev,A}>\lambda_{ev,B}$; blue regions) or larger $p$ value ($0>\lambda_{ev,B}>\lambda_{ev,A}$; red regions), depending on which of the two stable exponents dominates. There are also domains where the unbalanced two-cluster states are unstable ($\lambda_{ev,A/B}>0$, shown gray) and where only the synchronous stationary state is stable (white). For convenience, each case is assigned with a discrete variable $d\in\{1,2,3,4\}$. Evolution of cluster partition under noise, described by variation of partition parameter $p(t)$, is illustrated in Fig.~\ref{fig:ncl}(b)-(e) for typical solutions from several regions I-VI. While typical states from regions I and V eventually stabilize at a certain $p$ level close to initial one, displaying persistence under noise, a representative state from region II migrates to region III. Interestingly, the asymmetric leap-frog solution from region IV evolves toward balanced partition $p=1/2$, gaining antiphase space-time symmetry.

\begin{figure}[ht]
\vspace{-0.2cm}
\centering
\includegraphics[scale=0.3]{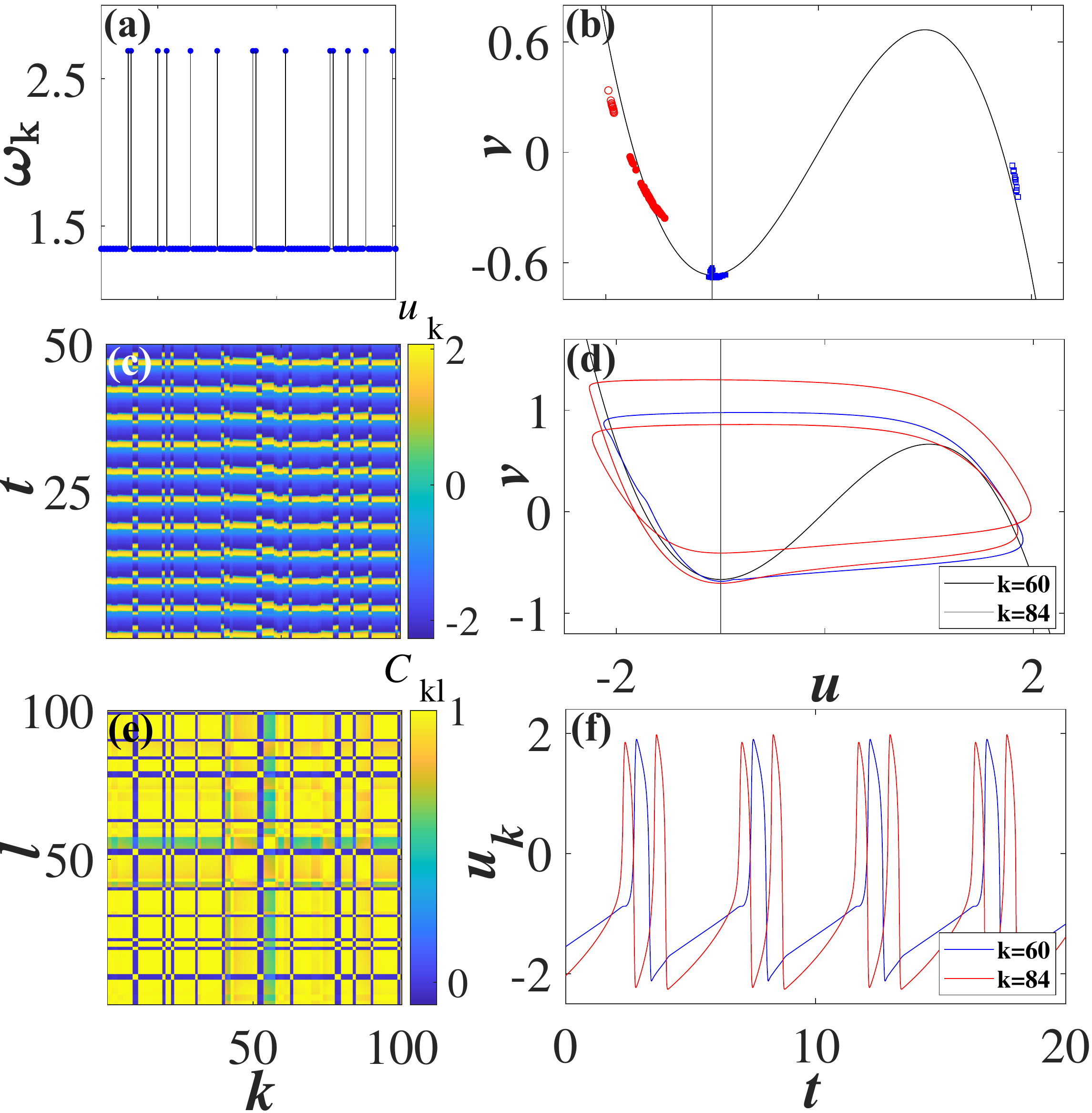}
\vspace{-0.3cm}
\caption{Solitary state SS1 in a non-locally coupled array ($N=100,\phi=1.85,r=0.2,\kappa=0.4,a=1.001,\varepsilon=0.05$). (a) Spatial profile of the
average local spiking frequencies $\omega_k$; (b) Red and blue: snapshots of local variables $(u_k,v_k)$ at two time moments $t_1<t_2$, black: nullclines of an isolated unit; (c) Spatiotemporal evolution of  activator variables $u_k(t)$; (d) Phase portraits $(u_k(t),v_k(t))$ for units representing solitary/minority ($k=84$) and synchronized/majority ($k=60$) cluster; (e) Cross-correlation matrix $C_{kl}$; (f) Time traces $u_k(t)$ for two representative units from (d).}\label{fig:ss1}
\vspace{-0.1cm}
\end{figure}
\paragraph{Solitary states in non-locally coupled arrays} We show that the prevalent solitary states in non-locally coupled arrays, SS1 and SS2, inherit their dynamical features from the corresponding unbalanced cluster states from Fig.~\ref{fig:std}(a). In particular, state SS1, illustrated in Fig.~\ref{fig:ss1}, is a dynamical counterpart of the two-cluster state from region I, whereas SS2 (not shown) derives from the cluster state from region V. In both cases, solitary states occur within the same $\phi$ interval and preserve frequency locking of the cluster states, but due to nonlocal interactions and the associated fluctuations of local mean-fields, the clusters of solitary and typical units are fuzzy \cite{JMK15} rather than exact. Spatial profile of average local frequencies $\omega_k=2\pi M_k/\Delta$, where $M_k$ is the number of complete rotations around the origin within the time interval $\Delta$, indeed shows a 2:1 frequency ratio between solitary and typical units. Analogy to two-cluster state in terms of local phase portraits and time traces $u_k(t)$ is illustrated in Fig.~\ref{fig:ss1}(d) and (f). Intrinsic dynamics of SS1 is characterized by the cross-correlation matrix $C_{kl}=\frac{\langle \hat{u}_k(t)\hat{u}_l(t) \rangle_T}{\sqrt{\langle \hat{u}_k(t)^2 \rangle_T\langle \hat{u}_l(t)^2 \rangle_T}}$, where $\langle \cdot \rangle_T$ refers to time averaging, while  $\hat{u}_k(t)=u_k(t)-\langle u_k(t)\rangle_T$ are the deviations of local activator variables from their means, cf. Fig.~\ref{fig:ss1}(e).

\begin{figure}[ht]
\centering
\includegraphics[scale=0.3]{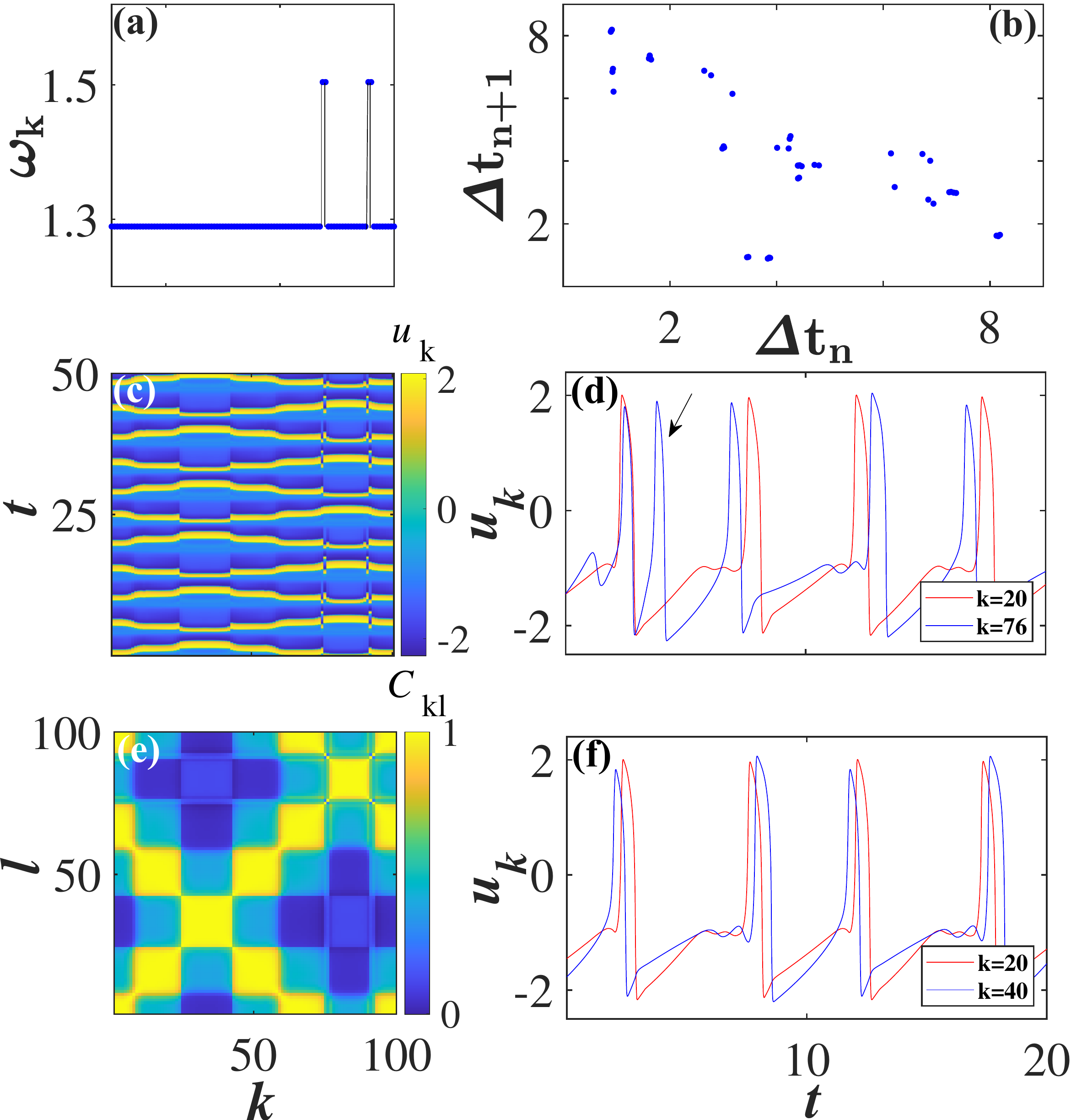}
\vspace{-0.3cm}
\caption{Chaotic solitary state SS3 ($N=100,\phi=1.788,r=0.2,\kappa=0.4,a=1.001,\varepsilon=0.05$)
(a) Spatial profile of $\omega_k$; (b) First return map of interspike intervals
$\Delta t_{n+1}(\Delta t_n)$ for a solitary unit $k=76$; (c) Space-time evolution of $u_k(t)$; (d) Time traces $u_k(t)$ for a solitary unit ($k=76$) and a unit from synchronized cluster ($k=20$); (e) Cross-correlation matrix $C_{kl}$; (f) Time traces $u_k(t)$ show leap-frog dynamics within synchronized cluster ($k=20$ and $k=40$).}\label{fig:ss3}
\vspace{-0.2cm}
\end{figure}
We also find less likely solitary states that do not have two-cluster state counterparts.
A typical example is a state SS3, see Fig.~\ref{fig:ss3}, which unlike SS1 and SS2, features leap-frog dynamics, both for solitary-typical pairs of units, and pairs of only solitary or typical units, cf. Fig.~\ref{fig:ss3}(d) and (f). SS3 admits chaotic rather than periodic dynamics, see Fig.~\ref{fig:ss3}(b), and emerges due to nonlocal interactions, which induce self-localized excitations \cite{WOS15} at interfaces separating domains with distinct dynamics. Frequency profile $\omega_k$ shows two clusters, but with a frequency ratio distinct from SS1, cf. Fig.~\ref{fig:ss1}(a). The difference in average frequencies here derives from random events where solitary units emit two successive spikes instead of a single spike followed by a subthreshold oscillation, see the arrow in Fig.~\ref{fig:ss3}(d). Also note the more complex correlation structure of SS3 compared to SS1, cf. Fig.~\ref{fig:ss3}(e) and Fig.~\ref{fig:ss1}(e).
\begin{figure}[ht]
\vspace{-0.3cm}
\centering
\includegraphics[scale=0.19]{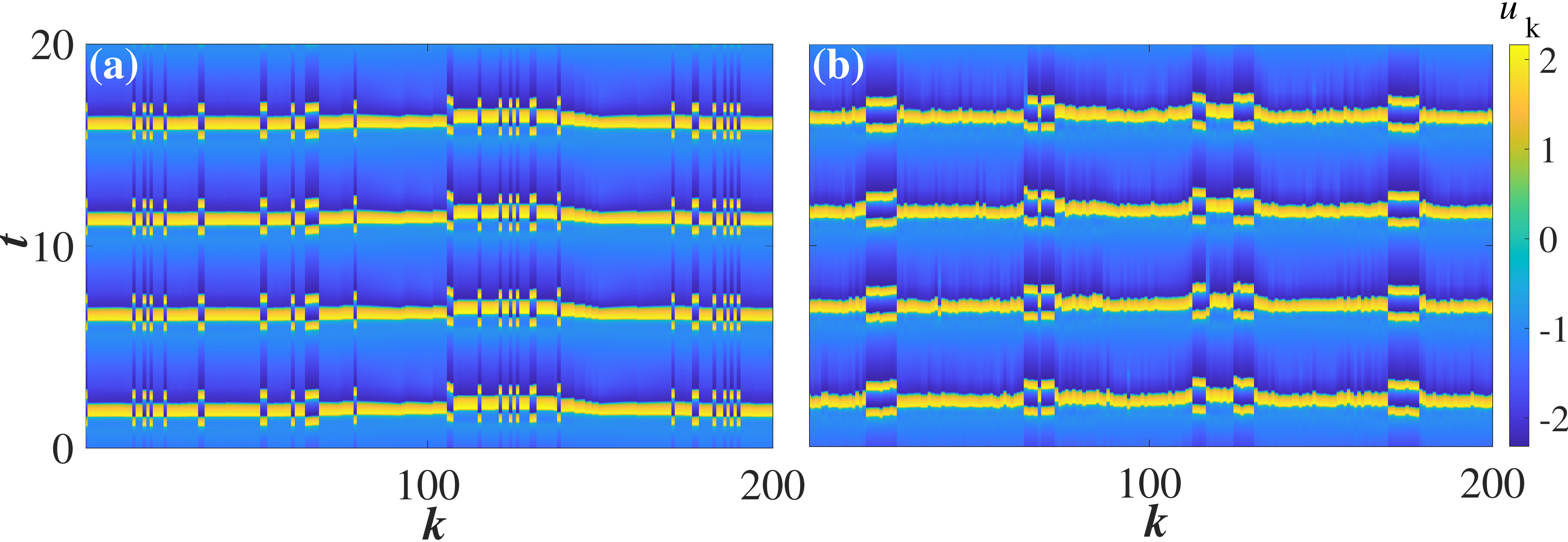}
\vspace{-0.4cm}
\caption{Transformation of solitary states into patched states under noise. (a) Typical spatiotemporal dynamics before introduction of noise (SS1 state); (b) Patched pattern developed from SS1 under noise $\sigma=0.0011$. Time axes in (a) and (b) are shifted. Parameters are
$N=200,\phi=2.0,r=0.2,\kappa=0.4,a=1.001,\varepsilon=0.05$.}\label{fig:patches}
\vspace{-0.4cm}
\end{figure}

While in \emph{locally} coupled excitable systems noise may strongly influence pattern formation by inducing, enhancing or controlling wave propagation, spiral dynamics and pacemaking \cite{LGNS04}, here the deterministic dynamics of a nonlocally coupled array features an additional landmark, namely extensive multistability, such that the main impact of noise is qualitatively different. We find that the noise reduces the system’s multistability, suppressing the solitary states. This reflects the effect called noise-induced preference of attractors \cite{PF14}, which may be understood as follows: in a highly multistable system, stability boundaries of attractors become smeared by noise, and only those with sufficiently large basins of attraction remain visible. For small noise, unbalanced splitting into frequency clusters is still preserved, but the preferred spatial distribution of minority units is \emph{localized} rather than random. This gives rise to \emph{patched} patterns with 1:2 subharmonic frequency locking. A typical example is provided in Fig.~\ref{fig:patches}, which shows how an initial configuration corresponding to SS1 transforms under noise into a state of patched synchrony \cite{OOHS13}.

\paragraph{Disussion} We have demonstrated the mechanism of onset and the link between the unbalanced periodic two-cluster states and solitary states, as a form of weak chimeras, in coupled excitable systems where repulsion dominates over attractive interactions. The fact that the prevalent solitary states in non-locally coupled arrays inherit their dynamical features from unbalanced cluster states in globally coupled networks is to a certain extent qualitatively similar to the finding for globally coupled Stuart-Landau oscillators, where clustering is identified as a symmetry-breaking step required for emergence of chimeras \cite{SK15}. The peculiar property of symmetry-broken states we discovered is the leap-frog dynamics. Such a behavior derives from multiscale character of the system, and in particular the phase-sensitive excitability of relaxation oscillations, underlying their strong sensitivity to perturbations in vicinity of the canard transition \cite{FOW18,EFW19}. Current results, together with \cite{RW12,EFW19}, indicate the importance of this concept to pattern formation in multiscale systems, universal to coupled excitable units and coupled oscillators. Multistability of the system dynamics is suppressed by noise, as a hallmark of noise-induced preference of attractors, an effect previously corroborated in coupled oscillators \cite{K97}, the H\'{e}non map \cite{MP12} and multistable fiber lasers \cite{PJSHT11}. Small noise is found to influence pattern formation by promoting homogeneous patched patterns at the expense of solitary states. Given that solitary states in coupled oscillators may mediate the transition from complete synchrony to chimeras \cite{JMK15}, an interesting open question is whether they may play a similar role in coupled excitable systems.

\begin{acknowledgments}
I.F. acknowledges funding from Institute of Physics Belgrade through grant by Ministry of Education,
Science and Technological Development of Republic of Serbia. AZ acknowledges support from the Deutsche Forschungsgemeinschaft (DFG) within the framework of the SFB 910, Projektnummer 163436311.
\end{acknowledgments}


\begin{thebibliography}{10}

\bibitem{KB02}{Y. Kuramoto, and D. Battogtokh, Nonlinear Phenom. Complex Syst.
\textbf{5},380 (2002).}

\bibitem{AS04}{D. M. Abrams, and S. H. Strogatz, Phys. Rev. Lett. \textbf{93},
174102 (2004).}

\bibitem{PRK03}{A. Pikovsky, M. Rosenblum, and J. Kurths, \emph{Synchronization: A Universal
Concept in Nonlinear Sciences}(Cambridge University Press, Cambridge, 2003).}

\bibitem{M10}{A. E. Motter, Nat. Phys. \textbf{6}(3), 164 (2010).}

\bibitem{KHK19}{F. P. Kemeth, S. W. Haugland, K. Krischer, Chaos \textbf{29}, 023107 (2019).}

\bibitem{KGO15}{W. L. Ku, M. Girvan, and E. Ott, Chaos \textbf{25}, 123122 (2015).}

\bibitem{IA15}{A. Ismail and P. Ashwin, Dyn. Syst. \textbf{30}, 122 (2015).}

\bibitem{HMM93}{D. Hansel, G. Mato, and C. Meunier, Phys. Rev. E \textbf{48}, 3470 (1993).}

\bibitem{PA15}{M. J. Panaggio, and D. M. Abrams, Nonlinearity \textbf{28}, R67 (2015).}

\bibitem{O18}{O. E. Omel'chenko, Nonlinearity \textbf{31}, R121 (2018).}

\bibitem{Z20}{A. Zakharova, \emph{Chimera Patterns in Networks: Interplay Between Dynamics,
Structure, Noise, and Delay - Understanding Complex Systems}, (Springer Nature, Switzerland, 2020).}

\bibitem{PJAS21}{F. Parastesh, S. Jafari, H. Azarnoush, Z. Shahriari, Z. Wang,
S. Boccaletti, and M. Perc, Phys. Rep. \textbf{898}, 1 (2021).}

\bibitem{MPR14}{Y. Maistrenko, B. Penkovsky, and M. Rosenblum, Phys. Rev. \textbf{E} 89, 060901(R) (2014).}

\bibitem{MSO20}{V. Maistrenko, O. Sudakov, and O. Osiv, Chaos \textbf{30}, 063113 (2020).}

\bibitem{JMK15}{P. Jaros, Y. Maistrenko, and T. Kapitaniak, Phys. Rev. E \textbf{91}, 022907 (2015).}

\bibitem{RASZ19}{E. Rybalova, V.S. Anishchenko, G.I. Strelkova, and A. Zakharova, 
Chaos \textbf{29}, 071106 (2019).}

\bibitem{BPSY20}{R. Berner, A. Polanska, E. Schöll, and S. Yanchuk, Eur. Phys. J. Spec. Top. 
\textbf{229}, 2183 (2020).}

\bibitem{JBLDKM18}{P. Jaros, S. Brezetsky, R. Levchenko, D. Dudkowski, T. Kapitaniak,
and Y. Maistrenko, Chaos \textbf{28}, 011103 (2018).}

\bibitem{OMHS11}{I. Omelchenko, Y. Maistrenko, P. H\"{o}vel, and E. Sch\"{o}ll, Phys. Rev. Lett. \textbf{106}, 234102 (2011).}

\bibitem{AB15}{P. Ashwin, and O. Burylko, Chaos \textbf{25}, 013106 (2015).}

\bibitem{SK15}{L. Schmidt, and K. Krischer, Phys. Rev. Lett. \textbf{114}, 034101 (2015).}

\bibitem{ZM21}{Y. Zhang, and A. Motter, Phys. Rev. Lett. \textbf{126}, 094101 (2021).}

\bibitem{LGNS04}{B. Lindner, J. Garc\'{\i}a-Ojalvo, A. Neiman, and L. Schimansky-Geier,
Phys. Rep. \textbf{392} 321 (2004).}

\bibitem{I07}{E. M. Izhikevich, \emph{Dynamical Systems in Neuroscience: The Geometry
of Excitability and Bursting} (MIT Press, Cambridge, MA, 2007).}

\bibitem{BRZ15}{I. Belykh, R. Reimbayev, and K. Zhao, Phys. Rev. E \textbf{91}, 062919 (2015).}

\bibitem{EFW19}{S. R. Eydam, I. Franovi\'{c}, and M. Wolfrum, Phys. Rev. E \textbf{99},
042207 (2019).}

\bibitem{GE02}{P. Goel, and G. B. Ermentrout, Physica D (Amsterdam) \textbf{163}, 191 (2002).}

\bibitem{AKW03}{C. D. Acker, N. Kopell, and J. A. White, J. Comp. Neurosci. \textbf{15}, 71 (2003).}

\bibitem{OM09}{M. Oh, and V. Matveev,J. Comput. Neurosci. \textbf{26}, 303 (2009).}

\bibitem{FTVB12}{I. Franovi\'c, K. Todorovi\'c, N. Vasovi\'c, and N. Buri\'c,
Phys. Rev. Lett. \textbf{108}, 094101 (2012).}

\bibitem{ZSAS17}{A. Zakharova, N. Semenova, V. Anishchenko, and E. Sch\"{o}ll,
Chaos \textbf{27}, 114320 (2017).}

\bibitem{SZAS16}{N. Semenova, A. Zakharova, V. Anishchenko, and E. Sch\"{o}ll,
Phys. Rev. Lett. \textbf{117}, 014102 (2016).}

\bibitem{SEM20}{N.Semenova, Eur. Phys. J. Special Topics 229, 2295 (2020)}

\bibitem{PF14}{A. N. Pisarchik, and U. Feudel, Phys. Rep. \textbf{540}, 167 (2014).}

\bibitem{K97}{K. Kaneko, Phys. Rev. Lett. \textbf{78}, 2736 (1997).}

\bibitem{PJSHT11}{A. N. Pisarchik, R. Jaimes-Reátegui, R. Sevilla-Escoboza, G. Huerta-Cuillar,
and M. Taki, Phys. Rev. Lett. \textbf{107}, 274101 (2011).}

\bibitem{BE86}{S. M. Baer and T. Erneux, SIAM  J.  Appl.  Math. \textbf{46}, 721 (1986).}

\bibitem{MRJZ19}{M. Mikhaylenko, L. Ramlow, S. Jalan, and A. Zakharova, Chaos \textbf{29}, 
023122 (2019).}

\bibitem{SJMZ21}{L. Sch\"ulen, D. A. Janzen, E. S. Medeiros, and A. Zakharova, 
Chaos Soliton. Fract. \textbf{145}, 110670 (2021).}

\bibitem{RZS21}{E. V. Rybalova, A. Zakharova, and G.I. Strelkova, Chaos Soliton. 
Fract. \textbf{148}, 111011 (2021).}

\bibitem{OOHS13}{I. Omelchenko, O. E. Omel'chenko, P. H\"{o}vel, and E. Sch\"{o}ll,
Phys. Rev. Lett. \textbf{110}, 224101 (2013).}

\bibitem{DRL12}{A. Destexhe, and M. Rudolph-Lilith, \emph{Neuronal noise} (Springer, New York, 2012).}

\bibitem{RZ21}{R. Ronge, and M. A. Zaks, Phys. Rev. E \textbf{103}, 012206 (2021).}

\bibitem{ZT16}{M. A. Zaks, and P. Tomov, Phys. Rev. E \textbf{93}, 020201(R) (2016).}

\bibitem{PP16}{A. Pikovsky, and A. Politi, \emph{Lyapunov Exponents: A Tool to Explore Complex Dynamics}
(Cambridge University Press, Cambridge, 2016).}

\bibitem{PPM01}{A. Pikovsky, O. Popovych, and Y. Maistrenko, Phys. Rev. Lett. \textbf{87}, 044102 (2001).}

\bibitem{ZP17}{M. Zaks, and A. Pikovsky, Sci. Rep. \textbf{7}, 4648 (2017).}

\bibitem{AUTO12}{E. J. Doedel, and B. E. Oldeman, AUTO-07P: Continuation and bifurcation
software for ordinary differential equations, (Concordia University, Montreal, 2012).}

\bibitem{DGKKOW12}{M. Desroches, J. Guckenheimer, B. Krauskopf, C. Kuehn,
H. M. Osinga, and M. Wechselberger, SIAM Rev. \textbf{54}, 211 (2012).}

\bibitem{K15}{C. Kuehn, \emph{Multiple Time Scale Dynamics} (Springer International
Publishing, Switzerland, 2015).}

\bibitem{FOW18}{I. Franovi\'{c}, O. E. Omel'chenko, and M. Wolfrum,
Chaos \textbf{28}, 071105 (2018).}

\bibitem{WOS15}{M. Wolfrum, O. E. Omel'chenko, and J. Sieber, Chaos \textbf{25}, 053113 (2015).}

\bibitem{RW12}{H. G. Rotstein and H. Wu, Phys. Rev. E \textbf{86}, 066207 (2012).}


\bibitem{MP12}{B. E. Martínez-Z\'{e}rega, and A. N. Pisarchik, Commun. Nonlinear Sci. Numer.
Simul. \textbf{17}, 4023 (2012).}




\end{thebibliography}
\end{document}